\documentclass[12pt]{article}
\usepackage{a4wide}
\usepackage{amssymb}
\begin{document}
{\renewcommand{\thefootnote}{\fnsymbol{footnote}}
\begin{center}
{\LARGE  Effective constraint algebras with structure functions}\\
\vspace{1.5em}
Martin Bojowald\footnote{e-mail address: {\tt bojowald@gravity.psu.edu}}
and Suddhasattwa Brahma\footnote{e-mail address: {\tt sxb1012@psu.edu}}
\\
\vspace{0.5em}
Institute for Gravitation and the Cosmos,\\
The Pennsylvania State
University,\\
104 Davey Lab, University Park, PA 16802, USA\\
\vspace{1.5em}
\end{center}
}

\setcounter{footnote}{0}

\begin{abstract}
  This article presents the result that fluctuations and higher moments of a
  state do not imply quantum corrections in structure functions of constrained
  systems. Consequences for canonical quantum gravity, whose structure
  functions encode space-time structure, are discussed. In particular,
  deformed algebras found in models of loop quantum gravity provide reliable
  information even in the Planck regime.
\end{abstract}

\section{Introduction}

Canonical gravity provides examples for constraint algebras with structure
functions that depend on phase-space variables. With this feature, one cannot
refer to well-known Lie algebra representation theory in order to find
candidates for a quantization. Direct attempts to construct first-class
quantizations of the constraints encounter several difficulties related to the
anomaly problem and have, for the general theory of arbitrary gravitational
fields and space-time geometries, not been successful. (There has been some
progress in loop quantum gravity
\cite{AnoFree,QSDI,TwoPlusOneDef,TwoPlusOneDef2,AnoFreeWeak}, but no
anomaly-free off-shell algebra for the full theory is known yet.) Since
complications appear at many different levels, it is useful to split the
problem. Instead of (i) finding suitable Hilbert spaces (or other structures)
to represent the constraints, (ii) constructing such representations, (iii)
making sure that the resulting constraint operators obey a closed algebra and
are therefore first class and anomaly-free, and then (iv) solving the
constraint equations and trying to interpret physical states, we will in this
article focus on the algebraic part of the problem: the effect of quantum
corrections on possible algebraic relations between the constraints, amounting
to part (iii) in our list. As we will see, not only can this question be
addressed in isolation from the others; it also provides valuable insights
about the quantum systems analyzed.

The first two steps of finding suitable Hilbert spaces and representations of
the constraints can be circumvented by using the canonical effective
formalism, briefly described below. As we will find, effective constraints
(defined as expectation values of constraint operators in some family of
states) allow one to draw rather strong conclusions about possible (or
impossible) quantum constraint algebras even in regimes in which one does not
expect semiclassical states to present good approximations. Such results are
especially important in the context of possible Planckian quantum-gravity
effects such as signature change in models of loop quantum gravity
\cite{Action,SigChange,PhysicsToday}.

For a specific result on the form of constraint algebras, we refer to a
quantum system subject to finitely many constraints $\hat{C}_I$,
$I=1,\ldots,N$ obeying an algebra
\begin{equation} \label{Algebra}
 [\hat{C}_I,\hat{C}_J]=i\hbar \hat{f}_{IJ}^K\hat{C}_K
\end{equation}
with structure functions $f_{IJ}^K$, turned into operators
$\hat{f}_{IJ}^K$. We therefore assume that the corresponding classical system
can, in some way, be quantized without anomalies. Since $\hat{f}_{IJ}^K$ is an
operator, one could expect quantum corrections to modify the classical algebra
$\{C_I,C_J\}=f_{IJ}^K(x_i)C_K$ (with phase-space functions $f_{IK}^K(x_i)$
depending on variables $x_i$, $i=1,\ldots, n$) by fluctuation terms or other
moments of a state: If effective constraints are defined as
$\langle\hat{C}_I\rangle$, the expectation value taken in some states, their
algebra might contain the expectation value $\langle\hat{f}_{IJ}^K\rangle$
which, for $f_{IJ}^K(x_i)$ non-linear in $x_i$, differs from the classical
structure functions $f_{IJ}^K(\langle\hat{x}_i\rangle)$ by fluctuation terms
and higher moments.

If this were the case, the quantization would be anomaly-free because the
quantum and effective constraints would form closed systems. But the classical
algebra of constraints, and therefore the gauge transformations they generate,
would be deformed. In canonical gravity, the constraints generate gauge
transformations that are equivalent to space-time diffeomorphisms on the
solution space of the constraints. A deformation of the quantum or effective
algebra would mean that the underlying space-time structure could no longer be
given directly\footnote{There are examples in which a deformed constraint
  algebra can be ``undeformed'' to the classical algebra by a canonical
  transformation \cite{Absorb}. One can then introduce effective space-times
  of the classical type after a field redefinition. However, such space-time
  structures are not obtained for direct quantizations of the classical
  metric.} by a Riemannian manifold with an invariant line element.
Higher-curvature effective actions could not exist because the local form of
covariance would be modified and no longer be respected by the usual curvature
invariants. The form of curvature corrections is especially important in the
Planck regime. Studying properties of anomaly-free constraint algebras can
therefore provide useful clues for cosmological (and black-hole) scenarios.

Our main result will show that quantization, by itself, does not modify the
constraint algebra at an effective level, provided that the constraint
operators can directly be represented in an anomaly-free way and the structure
functions are free of ordering ambiguities. By a careful analysis of effective
constraints, as introduced in \cite{EffCons,EffConsRel}, we will find that
structure functions will not receive quantum corrections by fluctuations or
moments of a state, even if they are non-linear in phase-space variables. In
such a case, applied to gravity, the theory still has the standard space-time
structure and may be expressed by a higher-curvature effective
action. Deformations of the constraint algebra could appear only by
non-trivial regularization steps or modifications that are sometimes required
to introduce well-defined constraint operators, especially when methods of
loop quantum gravity are used. In this way, unexpected and potentially
significant new quantum space-time effects become possible. (See
\cite{ReviewEff} for a review of such effects.)

\section{Effective constraint algebras}

Our starting point is a quantum system with basic operators $\hat{x}_i$,
$i=1,\ldots n$, constrained by $N$ conditions that some operators $\hat{C}_I$,
$I=1,\ldots N$, annihilate physical states. We will not be concerned with the
specific form of the kernel of all $\hat{C}_I$ as a physical Hilbert space,
but only consider the important condition of off-shell closure of the algebra
(\ref{Algebra}), a pre-requisite for the existence of a well-defined physical
Hilbert space.

\subsection{Effective constraints}

We follow the definitions of effective constrained systems developed in
\cite{EffCons,EffConsRel}. For every constraint operator $\hat{C}_I$, we
introduce an infinite family of effective constraints
\begin{equation} \label{Cpol}
 C_{I,{\rm pol}}:= \langle\widehat{{\rm pol}} \hat{C}_I\rangle
\end{equation}
with $\widehat{\rm pol}$ an arbitrary polynomial in $\widehat{\Delta
  x_i}:=\hat{x}_i -\langle\hat{x}_i\rangle$, whose coefficients may be
functions of basic expectation values $\langle\hat{x}_i\rangle$. We do not fix
any state to compute these expectation values, but rather view expressions
such as $\langle\hat{x}_i\rangle$ or $\langle\hat{C}_I\rangle$ as functions on
the space of all states. Instead of using the Dirac quantization condition
$\hat{C}_I\psi=0$ for wave functions $\psi$, we impose infinitely many
conditions $C_{I,{\rm pol}}=0$ which require all expectation values and
moments of the constraints to vanish. The specific ordering chosen in
(\ref{Cpol}) guarantees that the effective constraints are first class if the
constraint operators are first class. However, since no symmetric ordering is
assumed, so that $\hat{C}_I$ always acts directly on the state, some effective
constraints in general take complex values. (See the appendix of
\cite{FluctEn} for more details on complex terms in effective constraints.)
We require reality (and positivity of the state as determined by uncertainty
relations \cite{Casimir,ClassMoments}) of expectation values and moments only
after the effective constrained system has been solved, amounting to a
transition from kinematical to physical states.

An infinite set of effective constraints is difficult to solve. It can,
approximately, be reduced to finite sets if we introduce, for all
$a_1+\cdots+a_n\geq 2$, the moments
\begin{equation}
 \Delta(x_1^{a_1}\cdots x_n^{a_n}):= \langle (\widehat{\Delta
   x_1})^{a_1}\cdots (\widehat{\Delta
   x_n})^{a_n}\rangle_{\rm Weyl}
\end{equation}
in totally symmetric (or Weyl) ordering. In a semiclassical state, the
$\hbar$ierarchy
\begin{equation} \label{Order}
 \Delta(x_1^{a_1}\cdots x_n^{a_n})= O(\hbar^{(a_1+\cdots+a_n)/2})
\end{equation}
allows us to expand the effective constraints by powers of $\hbar$. At any
fixed order, only a finite number of moments contributes, subject to a finite
number of effective constraints.

If we assume an operator such as a constraint $\hat{C}$ to be Weyl-ordered, we
can write its expectation value $\langle\hat{C}\rangle$ as a function of the
state parameterized by its basic expectation values and moments:
\begin{eqnarray}
 \langle\hat{C}\rangle &=& \langle C(\hat{x}_i)\rangle= \langle
 C(\langle\hat{x}_i\rangle+
 (\hat{x}_i-\langle\hat{x}_i\rangle))\rangle\nonumber \\
&=& C(\langle\hat{x}_i\rangle)+ \sum_{a_1,\ldots,a_n} \frac{1}{a_1!\cdots a_n!}
\frac{\partial^{a_1+\cdots+a_n} C(\langle\hat{x}_i\rangle)}{\partial
  \langle\hat{x}_1\rangle^{a_1}\cdots \partial\langle \hat{x}_n\rangle^{a_n}}
\Delta(x_1^{a_1}\cdots x_n^{a_n})\,. \label{Expand}
\end{eqnarray}
(This type of expansion is the same as used for Hamiltonians in unconstrained
systems, where it has been shown to produce correct low-energy effective
actions for anharmonic oscillators as well as higher time derivatives in an
adiabatic approximation \cite{EffAc,Karpacz,HigherTime}.)  If $\hat{C}$ is a
polynomial in basic operators, the expansion ends after a finite number of
terms and is exact; it simply rewrites a polynomial in basic operators
$\hat{x}_i$ as a polynomial in $\widehat{\Delta
  x_i}=\hat{x}_i-\langle\hat{x}_i\rangle$ with coefficients depending on
$\langle\hat{x}_i\rangle$. Otherwise, the expansion is formal and acquires the
meaning of an $\hbar$-expansion upon using (\ref{Order}). If $\hat{C}$ is a
polynomial in basic operators that is not Weyl-ordered or not even
symmetrically ordered, one can always write it as a sum of Weyl-ordered terms
(see for instance \cite{Casimir,Search}). If the ordering is not symmetric,
some coefficients of these Weyl-ordered terms will be complex. In what
follows, we will assume all constraints $\hat{C}_I$ to be polynomial in basic
operators, but do not restrict the ordering. Non-polynomial constraints would
give rise to formal power series in our statements.

Details of this expansion and the orders considered for different terms will
be important in what follows; see \cite{Casimir,Counting} for more details. A
moment of degree $a_1+\cdots+a_n$ is of the order $\hbar^{(a_1+\cdots+a_n)/2}$
according to (\ref{Order}), while expectation values of the basic operators
$\hat{x}_i$ are considered to be of order zero. However, when we impose
effective constraints we will be forced to mix these orders. For instance, if
$\hat{C}=\hat{p}^2$ is the square of a canonical momentum, the effective
constraint equation
$\langle\hat{C}\rangle=\langle\hat{p}\rangle^2+\Delta(p^2)=0$ identifies a
zeroth-order function $\langle\hat{p}\rangle^2$ with a first-order variable
$\Delta(p^2)=(\Delta p)^2$. For consistency, we therefore declare that the
classical constraints $C_I(\langle\hat{x}_i\rangle)=: C_I^{\rm class}$
evaluated in expectation values of basic operators are not of degree zero but
of degree two (or of the order of $\hbar$).

In order to be able to compute constraint algebras at the effective level, we
use a Poisson structure on the space of expectation values and moments. For
expectation values of some operators $\hat{A}$ and $\hat{B}$ (not necessarily
basic ones), 
\begin{equation} \label{Poisson}
 \{ \langle\hat{A}\rangle, \langle\hat{B}\rangle\}:= \frac{\langle
   [\hat{A},\hat{B}]\rangle}{i\hbar}
\end{equation}
is linear and satisfies the Jacobi identity. We extend it to arbitrary
functions of expectation values (such as the moments) by requiring the Leibniz
rule. Restricted to moments of some fixed maximum order, the Poisson structure
given by (\ref{Poisson}) is in general not symplectic. For canonical
$\hat{x}_i$, we have $\{\langle\hat{x}_i\rangle,\Delta(\cdots)\}=0$ for all
moments, and there is a closed but lengthy formula for Poisson brackets of
moments \cite{HigherMoments}. An important property of the latter is that the
Poisson bracket of a moment of order $m$ and one of order $n$ is of the order
$n+m-2$.

We can directly apply this Poisson bracket in order to derive useful
statements about effective constraint algebras. If we have (\ref{Algebra}), we
can easily derive
\begin{eqnarray*}
 [\widehat{\rm pol}_{\alpha}\hat{C}_I, \widehat{\rm pol}_{\beta}\hat{C}_J]
 &=& \Bigl(\widehat{\rm pol}_{\alpha} \widehat{\rm pol}_{\beta}
 \hat{f}_{IJ}^K + \widehat{\rm pol}_{\alpha} [\hat{C}_I, \widehat{\rm
   pol}_{\beta}] \delta_J^K+ \widehat{\rm pol}_{\beta} [\widehat{\rm
   pol}_{\alpha},\hat{C}_J] \delta_I^K\\
&&+ [\widehat{\rm pol}_{\alpha},
 \widehat{\rm pol}_{\beta}] \hat{C}_J\delta_I^K\Bigr) \hat{C}_K\\
&=:& \hat{F}_{I\alpha,J\beta}^K \hat{C}_K\,,
\end{eqnarray*}
referring to some numbering of polynomials indicated by Greek indices.
This calculation, using standard relationships of commutators, implies that
the effective constraints obey an algebra
\begin{equation} \label{EffAlgebra}
 \{C_{I,{\rm pol}_{\alpha}}, C_{J,{\rm pol}_{\beta}}\}= \sum_K
 C_{K,F_{I\alpha,J\beta}^K}
\end{equation}
which is not only closed, but (formally) free of structure functions. (We
obtain $F_{I\alpha,J\beta}^K$ as a polynomial in $\widehat{\Delta x_i}$ by
inserting $\hat{x}_i= \widehat{\Delta x_i}+ \langle\hat{x}_i\rangle$ in
$\hat{F}_{I\alpha,J\beta}^K$.) The former structure functions rather appear as
effective constraints (\ref{Cpol}) with higher-order polynomial
coefficients. (Similarly, one can formally eliminate structure functions in
operator algebras (\ref{Algebra}) by extending the original constrained system
$\{\hat{C}_I\}$ by new constraints
\begin{eqnarray}
\hat{C}_{IJ}&:=&[\hat{C}_I,\hat{C}_J]= \hat{f}_{IJ}^K\hat{C}_K\\
\hat{C}_{LIJ}&:=&[\hat{C}_L, \hat{f}_{IJ}^K\hat{C}_K]\\
\hat{C}_{MLIJ}&:=&[\hat{C}_M,\hat{C}_{LIJ}]\\
 &\cdots&\nonumber\\
 \hat{C}_{NM\cdots} &:=& [\hat{C}_N,\hat{C}_{M\cdots}]\,.
\end{eqnarray}
In general, this procedure yields infinitely many constraint operators,
spanning a subset of the left-ideal generated by $\hat{C}_I$ in the algebra of
operators on a kinematical Hilbert space. The left-ideal corresponds to the
set of all constraints $C_{I,{\rm pol}}$, while $\hat{C}_{IJ\cdots}$ are those
whose expectation values appear in the expansion (\ref{EffAlgebra}).)

\subsection{Closure}
\label{s:Closure}

The latter result is only formal and does not help very much for practical
purposes. Instead, we use (\ref{EffAlgebra}) to conclude, following
\cite{EffCons}, that a closed (first-class) algebra of quantum constraints
implies a closed (first-class) algebra of effective constraints. Moreover, if
one expands all parts of (\ref{EffAlgebra}) up to some order in $\hbar$ or in
moments, the reduced effective constraints obey a closed algebra to within
this order: As noted after (\ref{Poisson}), the Poisson bracket of two moments
of orders $m$ and $n$, respectively, is of the order $m+n-2\geq {\rm
  min}(m,n)$. Taking Poisson brackets therefore commutes with truncating by
orders of $\hbar$.

Effective constraint algebras can provide conclusions about possible forms of
quantum constraint algebras in models in which the latter are difficult to
find. For instance, if one can show that it is not possible for certain types
of effective constraints to close up to some order in $\hbar$, a closed set of
constraint operators quantizing the same system cannot exist. For general
statements, one may introduce suitable parameters in effective constraints to
take into account quantization ambiguities, such as factor ordering. Some of
these choices could then be ruled out if they were to lead to anomalous
effective algebras to some order. A second set of results can be found when
effective constraints of a certain type can be anomaly-free, but only if the
constraint algebra is deformed compared to the classical one. One can then
conclude that the quantum constraint algebra must be deformed as well. A large
class of such examples exists in loop quantum gravity with holonomy or
inverse-triad corrections
\cite{ConstraintAlgebra,LTBII,JR,ModCollapse,ScalarHol,ScalarHolInv,HigherSpatial}.

These conclusions about possible constraint algebras turn out to be rather
powerful. They would be valid even if the $\hbar$-orders used gave rise to
poor approximations of the dynamics generated on the solution space of the
constraints. Indeed, the conclusions we inferred are off-shell statements
independent of properties of the solution space of the constraints. In quantum
gravity, provided one studies a sufficiently large class of effective
constraints with parameterizations of all expected ambiguities, the form of
constraint algebras and the corresponding space-time structures can therefore
be considered reliable even in the Planck regime, in which an
$\hbar$-expansion of the dynamics may be doubtful.

\subsection{Moments do not deform constraint algebras}

By definition of the Poisson bracket (\ref{Poisson}) on state space, the
leading part of the effective constraint algebra is given by
$\{C_{I,1},C_{J,1}\}= \langle \hat{f}_{IJ}^K\hat{C}_K\rangle$ evaluated for
constant polynomials ${\rm pol}_{\alpha}= {\rm pol}_{\beta}=1$. The right-hand
side contains fluctuations and higher moments if it is expanded by
(\ref{Expand}). (Or rather, by applying (\ref{Expand}) after ordering the
product of $\hat{f}_{IJ}^K$ and $\hat{C}_K$ totally symmetrically in basic
operators. Even though the product in the given ordering may not even be
symmetric, it can, as noted, always be written as a sum of totally symmetric
terms some of which may have complex coefficients.)  If the expansion would
look like $\langle \hat{f}_{IJ}^K\hat{C}_K\rangle=
\langle\hat{f}_{IJ}^K\rangle \langle\hat{C}_K\rangle+\Delta(\cdots)$, where
$\Delta(\cdots)$ indicates any kind of moment term that vanishes when
effective constraints $C_{I,{\rm pol}}$ for non-constant ${\rm pol}$ are
satisfied, the classical structure functions would be subject to quantum
corrections in the effective algebra, applying (\ref{Expand}) to
$\langle\hat{f}_{IJ}^K\rangle$. However, such an expectation value appears in
the moment expansion only if the $\hat{C}_K$ are basic operators, in which
case there could not be structure functions.

Instead of producing corrections to the structure functions, the expansion
(\ref{Expand}) introduces higher-order effective constraints $C_{I,{\rm
    pol}}$: $\langle \hat{f}_{IJ}^K\hat{C}_K\rangle=\sum_K C_{K,f_{IJ}^K}$. In
order to make this series more explicit, we continue with two assumptions and
then discuss consequences depending on whether they are satisfied. For now, we
assume that:
\begin{enumerate}
\item[(i)] The $\hat{f}_{IJ}^K$ {\em faithfully quantize} the classical
  structure functions $f_{IJ}^K$, that is,
  $\hat{f}_{IJ}^K=f_{IJ}^K(\hat{x}_i)$ in some ordering of basic operators,
  using the classical structure functions $f_{IJ}^K(x_i)$.
\item[(ii)] The quantized structure functions $\hat{f}_{IJ}^K$ are
  {\em Weyl-ordered.}
\end{enumerate}
With these assumptions, we can, by (i), rewrite
\[
 \langle \hat{f}_{IJ}^K\hat{C}_K\rangle=
\langle f_{IJ}^K(\widehat{\Delta x_i}+\langle\hat{x}_i\rangle) 
\hat{C}_K\rangle 
\]
and, by (ii), expand
\begin{equation} \label{fExpand}
 f_{IJ}^K(\widehat{\Delta x_i}+\langle\hat{x}_i\rangle)=
 f_{IJ}^K(\langle\hat{x}_i\rangle)+  
 \sum_{a_1,\ldots a_n} \frac{1}{a_1!\cdots a_n!}
 \frac{\partial^{a_1+\cdots+a_n}f_{IJ}^K(\langle\hat{x}_i\rangle)}{\partial
 \langle\hat{x}_1\rangle^{a_1}\cdots
 \partial\langle\hat{x}_n\rangle^{a_n}} \left((\widehat{\Delta
 x_1})^{a_1}\cdots (\widehat{\Delta x_n})^{a_n}\right)_{\rm Weyl}
\end{equation}
as used in (\ref{Expand}). Thus,
\begin{equation} \label{fC} 
\{C_{I,1},C_{J,1}\}= \langle
  \hat{f}_{IJ}^K\hat{C}_K\rangle= f_{IJ}^K(\langle\hat{x}_i\rangle)
  C_{K,1}+ \sum_j \frac{\partial
    f_{IJ}^K(\langle\hat{x}_i\rangle)}{
 \partial\langle\hat{x}_j\rangle} C_{K,x_j}+\cdots
\end{equation}
with structure functions $f_{IJ}^K(\langle\hat{x}_i\rangle)$ multiplying
leading effective constraints $C_{K,1}=\langle\hat{C}_K\rangle$ and
higher-order constraints $C_{K,x_i}=\langle\widehat{\Delta
  x_i}\hat{C}_K\rangle$ with their own new structure functions depending on
$\langle\hat{x}_i\rangle$, and so on. The form of effective constraints and
moment expansions makes it clear that structure functions are not corrected by
quantum terms, provided our two assumptions of faithful and Weyl-ordered
structure functions are satisfied. Quantization then implies only the presence
of new kinematical degrees of freedom and constraints in the
algebra. Classical and quantum degrees of freedom (expectation values and
moments) may be mixed by gauge flows generated by the constraints, as shown by
new terms in (\ref{fC}). But the leading structure functions
$f_{IJ}^K(\langle\hat{x}_i\rangle)$ are left unmodified, and new structure
functions such as $\partial
f_{IJ}^K(\langle\hat{x}_i\rangle)/\partial\langle\hat{x}_j\rangle$ in
(\ref{fC}) do not depend on moments and are strictly related to the leading
structure functions $f_{IJ}^K(\langle\hat{x}_i\rangle)$.  This conclusion is
correct to all orders, no matter how far one expands in (\ref{fC}). Moments
may, however, appear in new structure functions for brackets that involve
higher-order constraints. For instance, using commutator relationships and
some of the methods of \cite{Counting}, we compute
\begin{eqnarray} \label{CCpol}
 \{C_{I,1},C_{J,{\rm pol}}\} &=& \frac{1}{i\hbar} \left(
   \langle[\hat{C}_I,\widehat{\rm pol}] \hat{C}_J\rangle+ \sum_i \left\langle
     \frac{\partial\widehat{\rm pol}}{\partial\langle\hat{x}_i\rangle}
     \hat{C}_J\right\rangle \langle[\hat{C}_I,\hat{x}_i]\rangle\right)+
   \langle\widehat{\rm pol}\hat{f}_{IJ}^K\hat{C}_K\rangle
\end{eqnarray}
and
\begin{eqnarray}\label{CxCx}
 \{C_{M,x_i},C_{N,x_j}\} &=& \frac{1}{i\hbar} \left(\langle[\widehat{\Delta
     x_i}\hat{C}_M, \widehat{\Delta x_j}\hat{C}_N]\rangle- 
   \langle[\hat{x}_i,\widehat{\Delta x_j}\hat{C}_N]\rangle C_{M,1}-
   \langle[\widehat{\Delta x_i}\hat{C}_M,\hat{x}_j]\rangle C_{N,1}\right)
 \nonumber\\
&&
 + C_{M,1}C_{N,1} \{\langle\hat{x}_i\rangle,\langle\hat{x}_j\rangle\}
\end{eqnarray}
for monomials ${\rm pol}$, and so on.  By expanding some of the remaining
coefficients and commutators using (\ref{fExpand}), one obtains new structure
functions with moment contributions. We will make this explicit in a further
analysis of (\ref{CCpol}) below.

Possible corrections to the classical structure functions can appear only in
two cases. First, if $\hat{f}_{IJ}^K$ cannot be obtained in Weyl ordering for
a first-class constraint algebra, that is if assumption (i) but not (ii) is
satisfied, there may be re-ordering terms in the expansion (\ref{fExpand})
which depend explicitly on $\hbar$. Such terms, evaluated for expectation
values, would be added to $f_{IJ}^K(\langle\hat{x}_i\rangle)$ in (\ref{fC})
and therefore present quantum corrections to the structure functions. However,
they do not come from moments but rather from a function depending on $\hbar$
and $\langle\hat{x}_i\rangle$.

Secondly, a deformed quantum algebra implies a deformed effective algebra. If
one has to regularize or modify the classical constraints in order to
represent them, assumption (i) is violated and the quantum structure functions
$\hat{f}_{IJ}^K=f_{IJ,{\rm reg}}^K(\hat{x}_i)$ are not directly obtained by
inserting basic operators in the classical structure functions. Accordingly,
the expansion in (\ref{fC}) gives leading structure functions $f_{IJ,{\rm
    reg}}^K(\langle\hat{x}_i\rangle) \not= f_{IJ}^K(\langle\hat{x}_i\rangle)$
not identical with the classical ones. (But still, the deformed structure
functions of effective constraints $C_{I,1}$ do not depend on moments.) This
is the case of interest in loop quantum gravity, which we will discuss in more
detail in the conclusions.

\subsection{The classical algebra within the effective system}

As noted, there are no quantum corrections in the constraint algebra
(\ref{fC}), but only new constraints for the quantum degrees of freedom. The
classical and undeformed constraint algebra could be extracted from (\ref{fC})
if there were consistent partial solutions of the constrained system, solving
all constraints in the set
\begin{equation}
 {\cal C}:=\left\{C_{K,x_1^{i_1}\cdots x_n^{i_n}}: (i_1,\ldots,i_n)\not=(0,\cdots,0)
 \mbox{ such that }
\frac{\partial^{i_1+\cdots+i_n} f_{IJ}^K(\langle\hat{x}_i\rangle)}{\partial
\langle\hat{x}_1\rangle^{i_1}\cdots \partial\langle\hat{x}_n\rangle^{i_n}}
\not=0\right\}
\end{equation}
(or all constraints in a larger set containing ${\cal C}$ but none of the
$C_{I,1}$).  For the partially solved system, an undeformed algebra for the
$C_{I,1}$ would then be obtained as a subalgebra within the remaining
effective constraint algebra. This subalgebra would be an exact and
uncorrected version of the classical algebra, even though the generators
$C_{I,1}$ in general differ from the classical constraints by quantum
corrections.

However, it is not easy to find consistent partial solutions. For the set
${\cal C}$ to be solvable consistently, it must form an ideal in the full
constraint algebra. The remaining constraints are then well-defined on the
partially reduced phase space obtained by factoring out the kernel of the
pre-symplectic form given by the pull-pack of the original symplectic form to
the partial constraint surface. In terms of Poisson structures (which one has
to use if the system of moments is truncated), one can consistently factor out
the Hamiltonian flow generated by constraints in ${\cal C}$. A Poisson
structure is then defined by requiring the embedding $i$ of the partial
constraint surface to be a Poisson map: $\{i^*f,i^*g\}=i^*\{f,g\}$ for all
functions $f$ and $g$ on the full phase space. The left-hand side defines a
Poisson bracket for all functions that weakly commute with the constraints in
${\cal C}$: if $i^*f_1=i^*f_2$, $f_2=f_1+C_f$ with some function $C_f$ that
vanishes on the partial constraint surface, and similarly $g_2=g_1+C_g$ if
$i^*g_1=i^*g_2$. The Poisson bracket is well-defined if
$i^*\{f_1,g_1\}=i^*\{f_2,g_2\}$, that is if $i^*\{C_f,g\}=0$ and
$i^*\{C_f,C_g\}=0$ for all $C_f$ and $C_g$ that vanish on the partial
constraint surface. For the latter condition, the constraints in ${\cal C}$
must form a subalgebra. For the former, applied to the unsolved constraints
which should have consistent Poisson brackets with one another, the
constraints in ${\cal C}$ must form an ideal.

In order to explore the conditions on an ideal, we evaluate the relations
(\ref{CCpol}) and (\ref{CxCx}) in more detail. We begin with (\ref{CCpol}) for
a monomial $\widehat{\rm pol}=\widehat{\Delta x_i}$, in which case we have
\begin{equation}
 \{C_{I,1},C_{J,x_i}\} = \langle (\hat{K}_{Ii}-\langle\hat{K}_{Ii}\rangle)
 \hat{C}_J\rangle+ \langle \widehat{\Delta x_i}\hat{f}_{IJ}^K \hat{C}_K\rangle\,,
\end{equation}
defining $\hat{K}_{Ii}:=[\hat{C}_I,\hat{x}_i]/i\hbar$. In the last term, we
can expand $\hat{f}_{IJ}^K$ as in (\ref{fExpand}), showing that only
higher-order constraints contribute. However, the first term may contribute
constraints $C_{J,1}$: We expand $\hat{K}_{Ii}$ as in (\ref{fExpand}) and,
denoting by a weak equality an equation up to higher-order constraints, obtain
\begin{equation} \label{CIJi}
 \{C_{I,1},C_{J,x_i}\} \approx - \sum_{\sum j_k\geq 2} \frac{1}{j_1!\cdots
   j_n!} \frac{\partial^{j_1+\cdots+j_n}
   K_{Ii}}{\partial\langle\hat{x}_1\rangle^{j_1}
   \cdots \partial\langle\hat{x}_n\rangle^{j_n}} \Delta(x_1^{j_1}\cdots
 x_n^{j_n}) C_{J,1}\,.
\end{equation}
The coefficients vanish for quadratic constraints, but in such a case there
could not be structure functions. In general, therefore, the higher-order
constraints cannot form an ideal. (Incidentally, (\ref{CIJi}) shows that
moments do appear in structure functions involving higher-order
constraints. However, these moments are not corrections to classical structure
functions since higher-order constraints have no classical analog.)  Moreover,
the relation (\ref{CxCx}) implies that higher-order constraints in general do
not even form a subalgebra: Even if we restrict the $x_i$ in (\ref{CxCx}) to
those that appear in structure functions, and even if we assume that these
$x_i$ all commute with one another, the last terms in (\ref{CxCx}) may be
non-zero but proportional to the constraints $C_{I,1}$ which are not contained
in ${\cal C}$.

Since it is, in general, not expected to have the rather strong condition of
an ideal satisfied, it is not always easy to reduce the effective constraint
algebra to one that exactly equals the classical one. If ${\cal C}$ does not
form an ideal but a sub-algebra, there is a well-defined Poisson structure on
the partial solution space, but Poisson brackets of some of the remaining
constraints depend on the embedding $i$ and not just on the partial constraint
surface as a subspace. In such a case, there is no unique reduced constraint
algebra, but one can find different realizations depending on the specific
embedding. We will illustrate this possibility in more detail by an example in
the following section.

\section{Example}

As an example, we consider a totally constrained system with two canonical
pairs $(q_1,p_1;q_2,p_2)$ as basic operators and two constraints $C_{1,{\rm
    class}}= q_1+q_2^3$, $C_{2,{\rm class}}= q_1q_2^2p_1+ \frac{1}{3}
q_2^3p_2$. Since we are interested in the off-shell algebra, a totally
constrained system has non-trivial properties.

It is easy to see that the operators $\hat{C}_1=\hat{q}_1+\hat{q}_2^3$ and
$\hat{C}_2= \frac{1}{2}(\hat{q}_1\hat{p}_1+ \hat{p}_1\hat{q}_1)
\hat{q}_2^2+ \frac{1}{3} \left(\hat{q}_2^3\hat{p}_2\right)_{\rm
  Weyl}$ provide a first-class representation with structure functions in the
algebra
\begin{equation}
 [\hat{C}_1,\hat{C}_2]= i\hbar \hat{q}_2^2 \hat{C}_1\,.
\end{equation}
(The totally symmetric ordering of $\hat{C}_2$ is convenient but not
necessary.)  In order to see potential fluctuation terms in structure
functions, we should expand effective constraints at least to fourth order in
$\hbar$. In the present case, we can use exact expressions for the relevant
constraints.

We have
\begin{eqnarray}
 C_{1,1}&=& \langle\hat{C}_1\rangle= \langle\hat{q}_1\rangle
 +\langle\hat{q}_2\rangle^3+ 3\langle\hat{q}_2\rangle \Delta(q_2^2)+
 \Delta(q_2^3)\\ 
C_{2,1} &=& \langle\hat{C}_2\rangle= \langle\hat{q}_1\rangle
\langle\hat{q}_2\rangle^2 \langle\hat{p}_1\rangle+ \frac{1}{3}
\langle\hat{q}_2\rangle^3 \langle\hat{p}_2\rangle\nonumber\\
&&+ 2\langle\hat{q}_2\rangle
\langle\hat{p}_1\rangle \Delta(q_1q_2)+ 2\langle\hat{q}_1\rangle
\langle\hat{q}_2\rangle \Delta(q_2p_1)+ (\langle\hat{q}_1\rangle
\langle\hat{p}_1\rangle+ \langle\hat{q}_2\rangle \langle\hat{p}_2\rangle)
\Delta(q_2^2)\nonumber\\
&&+ \langle\hat{q}_2\rangle^2(\Delta(q_1p_1)+\Delta(q_2p_2))\nonumber\\
&&+ \langle\hat{p}_1\rangle \Delta(q_1q_2^2)+ 2\langle\hat{q}_2\rangle
\Delta(q_1q_2p_1)+ \langle\hat{q}_1\rangle \Delta(q_2^2p_1)+ \frac{1}{3}
\langle\hat{p}_2\rangle \Delta(q_2^3) 
 +\langle\hat{q}_2\rangle \Delta(q_2^2p_2)\nonumber\\
&&+ \Delta(q_1q_2^2p_1)+ \frac{1}{3}\Delta(q_2^3p_2)\,.
\end{eqnarray}

We will need several Poisson brackets of moments of order larger than two,
which can be found using the general formula given in \cite{HigherMoments}:
\begin{eqnarray}
 \{\Delta(q_2^2), \Delta(q_2^2p_2)\} &=& 2\Delta(q_2^3)\\
 \{\Delta(q_2^2), \Delta(q_2^3p_2)\} &=& 2\Delta(q_2^4)\\
\{\Delta(q_2^3), \Delta(q_2p_2)\} &=& 3\Delta(q_2^3)\\
\{\Delta(q_2^3), \Delta(q_2^2p_2)\} &=&
-3\Delta(q_2^2)^2+3\Delta(q_2^4)\\
\{\Delta(q_2^3), \Delta(q_2^3p_2)\} &=&
-3\Delta(q_2^2)\Delta(q_2^3)+3\Delta(q_2^5)\,.
\end{eqnarray}
The effective constraint algebra is then
\begin{eqnarray}
  \{C_{1,1},C_{2,1}\} &=& \langle\hat{q}_1\rangle \langle\hat{q}_2\rangle^2+
  \langle\hat{q}_2\rangle^5 +(10
  \langle\hat{q}_2\rangle^3+\langle\hat{q}_1\rangle)  \Delta(q_2^2)+ 2
  \langle\hat{q}_2\rangle \Delta(q_1q_2)\nonumber \\
 &&+\Delta(q_1q_2^2)+ 10 \langle\hat{q}_2\rangle^2 \Delta(q_2^3) + 5
 \langle\hat{q}_2\rangle \Delta(q_2^4)+ \Delta(q_2^5)\\
&=& \langle\hat{q}_2\rangle^2 C_{1,1}+ 2\langle\hat{q}_2\rangle C_{1,q_2} +
C_{1,q_2^2} \,.\label{C1C2}
\end{eqnarray}
We have used the effective constraints
\begin{equation}
 C_{1,q_2}= \langle\widehat{\Delta q_2} \hat{C}_1\rangle=
 \Delta(q_1q_2)+3\langle\hat{q}_2\rangle^2 \Delta(q_2^2)+
 3\langle\hat{q}_2\rangle \Delta(q_2^3)+ \Delta(q_2^4)
\end{equation}
and
\begin{eqnarray}
  C_{1,q_2^2}= \langle(\widehat{\Delta q_2})^2 \hat{C}_1\rangle&=&
  \left(\langle\hat{q}_1\rangle+ 
    \langle\hat{q}_2\rangle^3\right) \Delta(q_2^2)\\
  &&+\Delta(q_1q_2^2)+ 3\langle\hat{q}_2\rangle^2
  \Delta(q_2^3)+ 3\langle\hat{q}_2\rangle \Delta(q_2^4)+
  \Delta(q_2^5)\,.\nonumber 
\end{eqnarray}
This specific example agrees with the general expansion (\ref{fC}). In
particular, the quadratic quantum structure function $\hat{q}_2^2$ does not
give rise to fluctuation corrections in effective structure functions. If we
had tried to absorb some part of $C_{1,q_2^2}$ in $\Delta(q_2^2)C_{1,1}$, we
would have been left with unconstrained terms in
\begin{eqnarray}
 \{C_{1,1},C_{2,1}\} &=& \left(\langle\hat{q}_2\rangle^2+\Delta(q_2^2)\right)
 C_{1,1}+ 2\langle\hat{q}_2\rangle C_{1,q_2}\nonumber\\
&&+ 3\langle\hat{q}_2\rangle^2
 \Delta(q_2^3)+ \Delta(q_1q_2^2)+
 3\langle\hat{q}_2\rangle\left(\Delta(q_2^4)-
   \Delta(q_2^2)^2\right)\nonumber\\ 
 &&+
 \Delta(q_2^5)- \Delta(q_2^2) \Delta(q_2^3)\,.
\end{eqnarray}

The relation (\ref{C1C2}) does not have exactly the classical form because new
constraints $C_{1,q_2}$ and $C_{1,q_2^2}$ appear, as a consequence of moments
as new quantum degrees of freedom. The higher-order constraints $C_{1,q_2}$
and $C_{1,q_2^2}$, in this case, form a subalgebra but not an ideal:
$\{C_{2,1},C_{1,q_2}\}= -2\langle\hat{q}_2\rangle\Delta(q_2^2)C_{1,1}-
2\langle\hat{q}_2\rangle^2 C_{1,q_2}$. If partial solutions are used, the
Poisson brackets of reduced constraints depend on the embedding of the partial
constraint surface.

For the sake of brevity, we illustrate the resulting algebra up to first order
in $\hbar$ (or second-order moments), in which case $C_{1,q_2}$ is the only
constraint in (\ref{C1C2}) without a classical analog. We use this constraint
to eliminate the fluctuation $\Delta(q_2^2)$ from our constraints by replacing
it with $\Delta(q_2^2)= -\frac{1}{3}\langle\hat{q}_2\rangle^{-2}
\Delta(q_1q_2)$ in $C_{1,1}$ and $C_{2,1}$ and whenever it appears in Poisson
brackets:
\begin{eqnarray}
 C_{1,1}'&=& \langle\hat{C}_1\rangle|_{\Delta(q_2^2)=
   -\frac{1}{3}\langle\hat{q}_2\rangle^{-2} 
\Delta(q_1q_2)}= \langle\hat{q}_1\rangle
 +\langle\hat{q}_2\rangle^3-\frac{1}{\langle\hat{q}_2\rangle} \Delta(q_1q_2)\\
C_{2,1}' &=& \langle\hat{C}_2\rangle|_{\Delta(q_2^2)=
  -\frac{1}{3}\langle\hat{q}_2\rangle^{-2} 
\Delta(q_1q_2)}= \langle\hat{q}_1\rangle
\langle\hat{q}_2\rangle^2 \langle\hat{p}_1\rangle+ \frac{1}{3}
\langle\hat{q}_2\rangle^3 \langle\hat{p}_2\rangle\nonumber\\
&&+ \left(2\langle\hat{q}_2\rangle
\langle\hat{p}_1\rangle- \frac{\langle\hat{q}_1\rangle
\langle\hat{p}_1\rangle+ \langle\hat{q}_2\rangle
\langle\hat{p}_2\rangle}{3\langle\hat{q}_2\rangle^2}\right)
 \Delta(q_1q_2)\nonumber\\
&&+ 2\langle\hat{q}_1\rangle
\langle\hat{q}_2\rangle \Delta(q_2p_1)+
\langle\hat{q}_2\rangle^2(\Delta(q_1p_1)+\Delta(q_2p_2))\,.
\end{eqnarray}
These new constraints satisfy the algebra
\begin{eqnarray}
  \{C_{1,1}',C_{2,1}'\} &=& \langle\hat{q}_1\rangle \langle\hat{q}_2\rangle^2+
  \langle\hat{q}_2\rangle^5 + \frac{\langle\hat{q}_1\rangle-2
    \langle\hat{q}_2\rangle^3}{3 \langle\hat{q}_2\rangle^2} \Delta(q_1q_2)
  \nonumber\\ 
  &=& \langle\hat{q}_2\rangle^2 C_{1,1}'+
  \frac{\Delta(q_1q_2)}{3\langle\hat{q}_2\rangle^2}
  \left(C_{1,1}'+\frac{1}{\langle\hat{q}_2\rangle}
    \Delta(q_1q_2)\right)\nonumber\\ 
  &=& \langle\hat{q}_2\rangle^2 C_{1,1}'+ O(2)\,.
\end{eqnarray}
Within the orders considered, the algebra is identical with the classical one
even though the constraints $C_{1,1}'$ and $C_{2,1}'$ contain quantum
corrections. 


\section{Conclusions}

We have shown, by Eq.~(\ref{fC}), that quantum back-reaction by moments of a
state on expectation values does not lead to deformations of constraint
algebras in anomaly-free quantizations, even though the constraints themselves
receive quantum corrections. In the remainder of this article we discuss two
applications of our general result to canonical quantum gravity. For this, we
need to assume that one can extend our constructions to systems with
infinitely many constraints and degrees of freedom, or that the local dynamics
of quantum gravity can be expressed in terms of finitely many degrees of
freedom as for instance in some discrete versions. We also assume that
non-polynomial constraints can be dealt with by the methods of this article
using formal power series. Classically, there are then structure functions
which depend on the metric of a spatial slice within space-time, or on the
configuration variables but not on the momenta in a canonical formulation such
as \cite{ADM}.

\subsection{Quantum space-time structures}

The first consequence of our result is the presence of a state-independent
space-time structure. Structure functions which encode hypersurface
deformations in space-time \cite{Regained}, corresponding to our $C_{K,1}$, do
not depend on moments of a state, but only on expectation values which are in
one-to-one correspondence with the classical degrees of freedom. Just as
classical gravity has a Riemannian space-time structure irrespective of which
dynamical equations one solves for the metric, canonical quantum gravity gives
rise to a well-defined space-time structure with algebraic relations of
hypersurface-deformation generators irrespective of equations satisfied by
quantum states. (However, quantization always introduces new degrees of
freedom with independent gauge transformations that contribute to the
algebra. Their commutation relations may depend more sensitively on the
state.)

If the constrained system can be quantized faithfully, it has Weyl-ordered
structure functions because the $f_{IJ}^K$ of classical gravity depend only on
configuration variables. As shown here, faithfulness implies that there are no
modifications to the classical space-time structure and its
covariance. Effective actions should therefore be of higher-curvature form
with the standard notion of covariance and Riemannian space-time
structures. 

The higher-curvature form of quantum corrections makes use of an additional
assumption, namely that there is a local effective action obtained after a
valid derivative expansion. In terms of moments in canonical effective
descriptions, the analog of a derivative expansion is an adiabaticity
condition for the moments \cite{EffAc,Karpacz,HigherTime}. If moments do not
change adiabatically in a certain regime, the corrections they imply cannot be
expressed by higher-derivative terms in effective equations, and there is no
local effective action in terms of the classical fields.  However, a non-local
effective action can be rewritten as a local effective action with auxiliary
fields which, from a formal perspective, turn out to be just the moments, seen
as independent degrees of freedom. (In general, moments are far from being
just auxiliary degrees of freedom. Still, they can formally serve such a
purpose in an effective action.) In either case, the theory, local or
non-local, enjoys the classical space-time structure and gauge behavior if the
canonical constraint algebra is undeformed.

At present, it is, of course, unkown if canonical gravity can be quantized in
an anomaly-free way. Nevertheless, the conclusion is in agreement
with results from non-canonical derivations of effective actions of gravity
\cite{EffectiveGR,BurgessLivRev}. In this context, our results reconcile the
covariant and canonical view, and thereby provide support for the canonical
effective approach.

\subsection{Signature change}

In models of loop quantum gravity, an alternative canonical approach to
quantize gravity, deformed constraint algebras have been found in a variety of
systems and with different methods. Some of them have been derived for
effective constraints that do not include moments but other characteristic
quantum-geometry corrections expected from loop quantizations: inverse-triad
corrections \cite{ConstraintAlgebra,LTBII} and holonomy modifications
\cite{JR,ScalarHol,ScalarHolInv,HigherSpatial}. The deformed constraint
algebras they suggest appear to be rather universal \cite{ConsCosmo} and agree
also with results from operator calculations in $2+1$-dimensional models
\cite{ThreeDeform,TwoPlusOneDef,TwoPlusOneDef2,AnoFreeWeak}. Moment terms have
not been included yet in effective calculations of constraint algebras, but
our results show that they would not modify or eliminate the deformations. At
most, re-ordering terms could lead to $\hbar$-corrections of deformed
structure functions. (In the case of holonomy modifications, which according
to \cite{JR,ScalarHol,ScalarHolInv,HigherSpatial,ThreeDeform},
deform the classical structure functions by a factor depending on the momentum
of the spatial metric, the modified structure functions $f_{IJ,{\rm reg}}^K$
no longer depend just on configuration variables.) 

Such ordering terms may affect the details of concrete models based on
constraint algebras with deformed structure functions, especially in strong
quantum regimes. Nevertheless, our main result that moment-dependent quantum
back-reaction terms do not affect the leading structure functions has an
important consequence also in this context: In all consistent effective models
found so far in the presence of holonomy corrections, the deformed structure
functions change sign around any local maximum of holonomies as functions of
the connection or extrinsic curvature. (The structure functions are
proportional to the second derivative of holonomy modification functions
\cite{JR,HigherSpatial,ConsCosmo}.) This change of sign can be interpreted as
signature change, with a quantum version of 4-dimensional Euclidean space
replacing Lorentzian space-time when the structure functions take the opposite
sign \cite{Action,SigChange,PhysicsToday}. Even if holonomy modification
functions are subjected to $\hbar$-corrections from factor ordering terms,
this general conclusion about signature change remains unaltered. (The only
assumption in its derivation is that the classical quadratic dependence of the
Hamiltonian constraint on the connection or extrinsic curvature is replaced by
some function of the basic expectation values, not necessarily of any specific
form such as a sine function often used in this context.) Our results
therefore show that moment terms do not affect the central statements about
signature change.

As we have discussed in Section~\ref{s:Closure}, results about effective
constraint algebras are reliable in regimes in which semiclassical
approximations of the dynamics may be expected to be poor. One can therefore
trust implications of deformed algebras even in the Planck regime. The main
such result is signature change at high curvature or density.

\section*{Acknowledgements}

This work was supported in part by NSF grant PHY-1307408. We thank Juan Reyes
for useful comments on a draft of this paper.


\begin{thebibliography}{10}

\bibitem{AnoFree}
T.\ Thiemann,
\newblock Anomaly-Free Formulation of Non-Perturbative,
  Four-Dimensional Lorentzian Quantum Gravity,
\newblock {\em Phys.\ Lett.\ B} 380 (1996) 257--264, [gr-qc/9606088]

\bibitem{QSDI}
T.\ Thiemann,
\newblock Quantum Spin Dynamics {(QSD)},
\newblock {\em Class.\ Quantum Grav.} 15 (1998) 839--873, [gr-qc/9606089]

\bibitem{TwoPlusOneDef}
A.\ Henderson, A.\ Laddha, and C.\ Tomlin,
\newblock Constraint algebra in LQG reloaded : Toy model of a ${\rm U}(1)^{3}$
  Gauge Theory I,
\newblock {\em Phys.\ Rev.\ D} 88 (2013) 044028, [arXiv:1204.0211]

\bibitem{TwoPlusOneDef2}
A.\ Henderson, A.\ Laddha, and C.\ Tomlin,
\newblock Constraint algebra in LQG reloaded : Toy model of an Abelian gauge
  theory - II Spatial Diffeomorphisms,
\newblock {\em Phys.\ Rev.\ D} 88 (2013) 044029, [arXiv:1210.3960]

\bibitem{AnoFreeWeak}
C.\ Tomlin and M.\ Varadarajan,
\newblock Towards an Anomaly-Free Quantum Dynamics for a Weak Coupling Limit of
  Euclidean Gravity,
\newblock {\em Phys.\ Rev.\ D} 87 (2013) 044039, [arXiv:1210.6869]

\bibitem{Action}
M.\ Bojowald and G.~M.\ Paily,
\newblock Deformed General Relativity and Effective Actions from Loop Quantum
  Gravity,
\newblock {\em Phys.\ Rev.\ D} 86 (2012) 104018, [arXiv:1112.1899]

\bibitem{SigChange}
J.\ Mielczarek,
\newblock Signature change in loop quantum cosmology, [arXiv:1207.4657]

\bibitem{PhysicsToday}
M.\ Bojowald,
\newblock Back to the beginning of quantum spacetime,
\newblock {\em Physics Today} 66 (2013) 35

\bibitem{Absorb}
R.\ Tibrewala,
\newblock Inhomogeneities, loop quantum gravity corrections, constraint algebra
  and general covariance,
\newblock {\em Class.\ Quantum Grav.} 31 (2014) 055010, [arXiv:1311.1297]

\bibitem{EffCons}
M.\ Bojowald, B.\ Sandh\"ofer, A.\ Skirzewski, and A.\ Tsobanjan,
\newblock Effective constraints for quantum systems,
\newblock {\em Rev.\ Math.\ Phys.} 21 (2009) 111--154, [arXiv:0804.3365]

\bibitem{EffConsRel}
M.\ Bojowald and A.\ Tsobanjan,
\newblock Effective constraints for relativistic quantum systems,
\newblock {\em Phys.\ Rev.\ D} 80 (2009) 125008, [arXiv:0906.1772]

\bibitem{ReviewEff}
M.\ Bojowald,
\newblock Quantum Cosmology: Effective Theory,
\newblock {\em Class.\ Quantum Grav.} 29 (2012) 213001, [arXiv:1209.3403]

\bibitem{FluctEn}
M.\ Bojowald,
\newblock Fluctuation energies in quantum cosmology,
\newblock {\em Phys.\ Rev.\ D} 89 (2014) 124031, [arXiv:1404.5284]

\bibitem{Casimir}
M.\ Bojowald and A.\ Tsobanjan,
\newblock Effective Casimir conditions and group coherent states,
\newblock {\em Class.\ Quantum Grav.} 31 (2014) 115006, [arXiv:1401.5352]

\bibitem{ClassMoments}
D.\ Brizuela,
\newblock A formalism based on statistical moments for classical and quantum
  dynamics: application to cosmology, [to appear]

\bibitem{EffAc}
M.\ Bojowald and A.\ Skirzewski,
\newblock Effective Equations of Motion for Quantum Systems,
\newblock {\em Rev.\ Math.\ Phys.} 18 (2006) 713--745, [math-ph/0511043]

\bibitem{Karpacz}
M.\ Bojowald and A.\ Skirzewski,
\newblock Quantum Gravity and Higher Curvature Actions,
\newblock {\em Int.\ J.\ Geom.\ Meth.\ Mod.\ Phys.} 4 (2007) 25--52,
  [hep-th/0606232]

\bibitem{HigherTime}
M.\ Bojowald, S.\ Brahma, and E.\ Nelson,
\newblock Higher time derivatives in effective equations of canonical quantum
  systems,
\newblock {\em Phys.\ Rev.\ D} 86 (2012) 105004, [arXiv:1208.1242]

\bibitem{Search}
M.\ Bojowald and D.~Simpson,
\newblock Factor ordering and large-volume dynamics in quantum cosmology,
  [arXiv:1403.6746]

\bibitem{Counting}
A.\ Tsobanjan,
\newblock Semiclassical states on Lie algebras, [in preparation]

\bibitem{HigherMoments}
M.\ Bojowald, D.\ Brizuela, H.~H.\ Hernandez, M.~J.\ Koop, and H.~A.\
  Morales-T\'ecotl,
\newblock High-order quantum back-reaction and quantum cosmology with a
  positive cosmological constant,
\newblock {\em Phys.\ Rev.\ D} 84 (2011) 043514, [arXiv:1011.3022]

\bibitem{ConstraintAlgebra}
M.\ Bojowald, G.\ Hossain, M.\ Kagan, and S.\ Shankaranarayanan,
\newblock Anomaly freedom in perturbative loop quantum gravity,
\newblock {\em Phys.\ Rev.\ D} 78 (2008) 063547, [arXiv:0806.3929]

\bibitem{LTBII}
M.\ Bojowald, J.~D.\ Reyes, and R.\ Tibrewala,
\newblock Non-marginal LTB-like models with inverse triad corrections from loop
  quantum gravity,
\newblock {\em Phys.\ Rev.\ D} 80 (2009) 084002, [arXiv:0906.4767]

\bibitem{JR}
J.~D.\ Reyes,
\newblock {\em Spherically Symmetric Loop Quantum Gravity: Connections to
  2-Dimensional Models and Applications to Gravitational Collapse},
\newblock PhD thesis, The Pennsylvania State University, 2009

\bibitem{ModCollapse}
A.\ Kreienbuehl, V.\ Husain, and S.~S.\ Seahra,
\newblock Modified general relativity as a model for quantum gravitational
  collapse,
\newblock {\em Class.\ Quantum Grav.} 29 (2012) 095008, [arXiv:1011.2381]

\bibitem{ScalarHol}
T.\ Cailleteau, J.\ Mielczarek, A.\ Barrau, and J.\ Grain,
\newblock Anomaly-free scalar perturbations with holonomy corrections in loop
  quantum cosmology,
\newblock {\em Class.\ Quant.\ Grav.} 29 (2012) 095010, [arXiv:1111.3535]

\bibitem{ScalarHolInv}
T.\ Cailleteau, L.\ Linsefors, and A.\ Barrau,
\newblock Anomaly-free perturbations with inverse-volume and holonomy
  corrections in Loop Quantum Cosmology, [arXiv:1307.5238]

\bibitem{HigherSpatial}
M.\ Bojowald, G.~M.\ Paily, and J.~D.\ Reyes,
\newblock Discreteness corrections and higher spatial derivatives in effective
  canonical quantum gravity, [arXiv:1402.5130]

\bibitem{ADM}
R.\ Arnowitt, S.\ Deser, and C.~W.\ Misner,
\newblock The Dynamics of General Relativity, In L.\ Witten, editor, {\em
  Gravitation: An Introduction to Current Research},
\newblock Wiley, New York, 1962.
\newblock Reprinted in R.\ Arnowitt, S.\ Deser, and C.~W.\ Misner,
\newblock The Dynamics of General Relativity,
\newblock {\em Gen.\ Rel.\ Grav.} 40 (2008) 1997--2027

\bibitem{Regained}
S.~A.\ Hojman, K.\ Kucha\v{r}, and C.\ Teitelboim,
\newblock Geometrodynamics Regained,
\newblock {\em Ann.\ Phys.\ (New York)} 96 (1976) 88--135

\bibitem{EffectiveGR}
J.~F.\ Donoghue,
\newblock General relativity as an effective field theory: The leading quantum
  corrections,
\newblock {\em Phys.\ Rev.\ D} 50 (1994) 3874--3888, [gr-qc/9405057]

\bibitem{BurgessLivRev}
C.~P.\ Burgess,
\newblock Quantum Gravity in Everyday Life: General Relativity as an Effective
  Field Theory,
\newblock {\em Living Rev.\ Relativity} 7 (2004), [gr-qc/0311082],
\newblock http://www.livingreviews.org/lrr-2004-5

\bibitem{ConsCosmo}
A.\ Barrau, M.\ Bojowald, G.\ Calcagni, J.\ Grain, and M.\ Kagan,
\newblock Anomaly-free cosmological perturbations in effective canonical
  quantum gravity, [arXiv:1404.1018]

\bibitem{ThreeDeform}
A.\ Perez and D.\ Pranzetti,
\newblock On the regularization of the constraints algebra of Quantum Gravity
  in $2+1$ dimensions with non-vanishing cosmological constant,
\newblock {\em Class.\ Quantum Grav.} 27 (2010) 145009, [arXiv:1001.3292]

\end{thebibliography}

\end{document}